\documentclass[conference]{IEEEtran}
\IEEEoverridecommandlockouts

\usepackage{cite}
\usepackage{amsmath,amssymb,amsfonts}
\usepackage{algorithmic}
\usepackage{graphicx}
\usepackage{textcomp}
\usepackage{xcolor}
\usepackage{booktabs}
\usepackage{bm}
\usepackage{subfigure}
\usepackage{multirow}
\def\BibTeX{{\rm B\kern-.05em{\sc i\kern-.025em b}\kern-.08em
    T\kern-.1667em\lower.7ex\hbox{E}\kern-.125emX}}
\begin{document}

\title{SAR: Self-Supervised Anti-Distortion Representation for End-To-End Speech Model\\
\thanks{*Corresponding author: Xulong Zhang, zhangxulong@ieee.org}
}

\author{
\IEEEauthorblockN{Jianzong Wang$^1$, Xulong Zhang$^{1,*}$, Haobin Tang$^{1,2}$, Aolan Sun$^1$, Ning Cheng$^1$, Jing Xiao$^1$}
\IEEEauthorblockA{$^1$\textit{Ping An Technology (Shenzhen) Co., Ltd.}
\\$^2$\textit{University of Science and Technology of China}
\\ https://largeaudiomodel.com}
% \textit{Shenzhen, China}
}

\maketitle

\begin{abstract}
In recent Text-to-Speech (TTS) systems, a neural vocoder often generates speech samples by solely conditioning on acoustic features predicted from an acoustic model. However, there are always distortions existing in the predicted acoustic features, compared to those of the groundtruth, especially in the common case of poor acoustic modeling due to low-quality training data. To overcome such limits, we propose a Self-supervised learning framework to learn an Anti-distortion acoustic Representation (SAR) to replace human-crafted acoustic features by introducing distortion prior to an auto-encoder pre-training process. The learned acoustic representation from the proposed framework is proved anti-distortion compared to the most commonly used mel-spectrogram through both objective and subjective evaluation.
\end{abstract}


\section{Introduction}
Text to speech synthesis (TTS)~\cite{kim2021conditional,zhao2022nnspeech,Tang2023Learning, zeng2020Prosody, zhang2022Semi} is a task of generation condition on speaker, each input text could map with many output speech. The contemporary TTS system is primarily comprised of three key modules: text analysis, acoustic model, and vocoder \cite{tan2021survey}. The text analysis module processes the input text for normalization and divides it into phonemes. The acoustic model is to build the mapping from phoneme embedding to acoustic features such as spectrum. The vocoder final transforms the acoustic feature into waveform for the synthesized audio.  
In the backend of a TTS system, generating acoustic features is achieved by utilizing an acoustic model. The resulting acoustic features are subsequently fed into a vocoder, which then produces speech samples. However, when the speech data used for training is of low quality, the predicted acoustic features often contain distortions or are incomplete, which will directly affect the synthesized speech quality. 

To overcome the above issues, we propose to make use of a high-level coherent structure that can be observed both across frequency and across time in speech signals. We believe that there exists a high-level structure that connects different acoustic parts and that speech can only be generated when the acoustic parts are coherent with one another. Moreover, with the observed correlations and the coherent structure, we may be able to infer the missing parts from other components if some of the components are missing. Similarly, when faced with corrupted acoustic features, the learned representation of the high-level coherent structure is believed to have the potential to help reconstruct undistorted speech. 
Besides, to make the auto-encoder focus more on representation learning, we only utilize recurrent structure in the encoder with its decoder as simple stacked fully-connected layers. The learned encoder without latent space masking is used as an acoustic feature extraction component. The newly extracted acoustic features are used to train an acoustic model. For the neural vocoder, for speech up training, we used a trained neural vocoder connected to the simple decoder in auto-encoder to do finetune training. During the vocoder training, we keep the latent space masking strategies to keep the anti-distortion property from overfitting.

To obtain this high-level coherent structure within speech, we turn to the methodologies of self-supervised learning. However, most of the speech representation learning approaches focus on learning contextual information that is more global and discard local details, and downstream tasks are mostly for classifications like speech recognition. Nonetheless, the local details are key factors in regression tasks such as speech synthesis. Thus, we propose a self-supervised representation learning method that encodes both the detailed local information and the contextual information, incorporated with a distortion-aware prior. For pre-training, an auto-encoder is built to reconstruct the mel-spectrogram, during which a distortion-aware prior is introduced by masking the latent space of the auto-encoder randomly with various ratios. The distortion-aware prior will force the auto-encoder to learn a high-level coherent structure capable of retrieving missing information from the rest of the latent space features. The learned latent space features are used to replace the human-crafted mel-spectrogram to build the downstream speech synthesis system. In our experiments, we adopt Tacotron2~\cite{shen_2017_natural} as the acoustic model and Waveglow\cite{prenger_2018_Waveglow} as the neural vocoder for TTS systems. 
% For objective evaluation, the Extended Short-Time Objective Intelligibility (ESTOI) scores are computed from copy-synthesized speech samples generated from the neural vocoders based on mel-spectrogram and the learned representations respectively. The input mel-spectrogram and the learned representation are corrupted on purpose by adding white noise and masking different levels. Subjectively, some Mean Opinion Score (MOS) tests are carried out on the TTS system built from low-quality training data. Both evaluation results show that the learned representation is superior in terms of anti-distortion capability.
The primary contributions of our work:
\begin{itemize}
    \item Distortion-aware priors are introduced into representation learning through a masking strategy to impart anti-distortion properties to the learned acoustic representations.
    \item The learned acoustic representations with masking can be used to construct TTS systems in place of hand-crafted mel-spectrograms to build more refined acoustic features.
    \item The joint training of vocoder and encoder help train TTS models with low-quality speech data.   
    
\end{itemize}

\section{Related Work}

Two main issues keep the acoustic features from being distortion-free. Firstly, although the recently proposed end-to-end acoustic models~\cite{wang_2017_tacotron,shen_2017_natural, lancucki2021fastpitch, li2019neural, yu2020, yi2019, yi2021, zeng2020aligntts,Tang2023QI-TTS} and neural vocoders~\cite{oord_2016_wavenet,prenger_2018_Waveglow,kalchbrenner_2018_efficient, matsubara2021full, oord2018parallel, kalchbrenner2018efficient}, compared to the conventional methods as in~\cite{ze_2013_statistical}, have shown great progress in synthesizing human-like speech, there is always a gap between the ground-truth acoustic features and the predicted ones. As the ground-truth acoustic features are used to train the acoustic model and the neural vocoder separately, the prediction error or the distortion always exists. Some researchers approach this issue by training neural vocoders with the predicted acoustic features of the counterpart acoustic model~\cite{shen_2017_natural}, but the method has its limits since it does not generalize to different acoustic models, and one needs to retrain neural vocoder each time when there is a change in the corresponding acoustic model. Secondly, the prediction capability of an acoustic model is affected by multiple factors such as the amount and the quality of speech data in the training set. Specifically, due to the great cost of collecting high-quality studio speech data, many data collection tasks are conducted in an unprofessional room with background noise and even reverberations. Meanwhile, some acoustic models may be trained with data of low sampling rate, due to the limitation of the codec on the device or the transmission channel. The low-quality speech data lead to bad convergence of acoustic models, and hence the distortions in predicted acoustic features occur.

In the frequency dimension, the fundamental frequency~\cite{morise_2016_world} and harmonics generated through the vibration of vocal cords are highly correlated mathematically, with each harmonic being a certain number of times of the fundamental frequency. Meanwhile, on the time scale, the dynamic features like delta and delta-delta cepstral coefficients adopted in Maximum Likelihood Parameter Generation (MLPG)~\cite{tokuda_2000_speech}  to help better predict static acoustic features indicate the existence of a contextual correlation between speech frames.

Speech representations are learned through auto-encoder in several previous works. Chorowski \textit{et al.}~\cite{chorowski_2019_unsupervised} exploited the auto-encoder to reconstruct spectrogram frames by applying different constraints to latent space. However, their work focused more on unsupervised speech-tokens mapping instead of speech synthesis. The auto-encoder in~\cite{takaki_2016_a} learns deep latent features for speech synthesis but no contextual information is incorporated. More recently, He \textit{et al.}~\cite{he_2019_rawnet} proposed an end-to-end structure to learn new acoustic features to replace human-crafted ones for neural vocoders to reduce inference costs while keeping high voice quality. It is hard to model in the time domain, and they have applied many tricks to make the waveform-to-waveform mapping trainable. Therefore, the representation learning in our work is performed in the frequency domain, with our focus being on training an anti-distortion acoustic representation for speech synthesis at the same time.  The idea to learn an intermediate feature similar to the mel-spectrogram is also inspired from~\cite{dunbar2017zero, versteegh2015zero}.

As for the self-supervised approaches, most of the self-supervised learning methods focus on utilizing a large amount of unlabeled data which are much easier to collect for representation pre-training and use the pre-trained representation encoder for downstream tasks. Bidirectional Encoder Representations from Transformers (BERT)~\cite{oord_2018_representation, devlin-etal-2019-bert}, which is pre-trained to learn informative semantic representation, is now widely used as a pre-trained model for almost all kinds of natural language processing tasks. In the field of speech, regarding the prediction of future information, Contrastive Predictive Coding~\cite{jiang_2019_improving} is one of the self-supervised learning methods to extract useful speech representation by predicting future data samples. In \cite{liu_2020_tera} speech representations are learned by reconstruction of altered frames of acoustic features. 

\section{PROPOSED METHOD}

\subsection{Self-supervised pre-training}

\begin{figure}[!htbp] 
    \centering 
    \includegraphics[width=\linewidth]{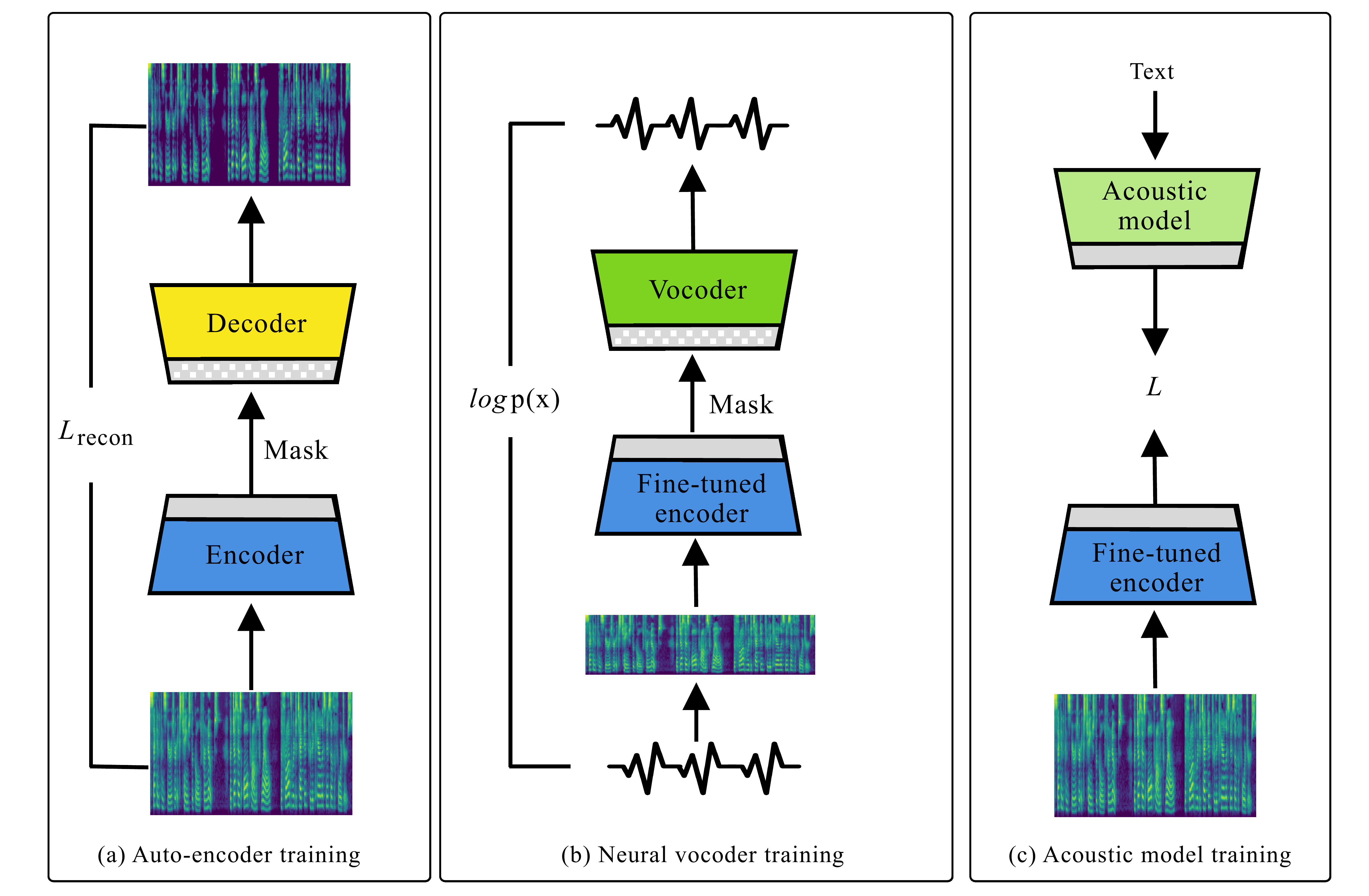}
    \caption{The architecture of speech representation learning. (a) The architecture of self-supervised auto-encoder. The corrupted blank in the blocks on the bottom of the decoder denotes masking on the encoder outputs, which are normally random dropouts. (b) Fine-tuning of the recurrent encoder during neural vocoder training. Similarly, the corrupted blocks below the vocoder denote masking blocks. 
    (c) Training of the Acoustic model using learned anti-distortion representations.}
    \label{imag1} 
\end{figure}

\subsubsection{Introducing distortion-aware prior through masking}
Self-supervised learning methods usually design an objective in order to learn informative representation from unlabeled data~\cite{liu_2020_tera,pennington_2014_glove,chorowski_2019_unsupervised, jiang_2020_speech}. Reconstruction is one of such objectives to extract representation by reconstructing unlabeled data through the specific model structure like auto-encoder~\cite{chorowski_2019_unsupervised, jiang_2020_speech, tang2021TGAVC}. However, most of them focus on learning contextual features and discard local features. In our proposal, targeting at speech synthesis tasks and local feature recovery, we choose to mask learned latent space features. By masking latent space features, the auto-encoder is forced to learn a higher-level coherent structure from a relatively low-level input mel-spectrogram. And these learned latent space features are the ones we use to replace human-crafted mel-spectrogram for downstream speech synthesis tasks. Specifically, as depicted in Figure \ref{imag1} (a), the latent space features at each time step are randomly masked at a certain ratio $\alpha$ during the training while $\alpha=0$ during the inference stage. The $\alpha$ is sampled from a uniform distribution in which the minimum and maximum are $0$ and ${\alpha}_{max}$ respectively. 

\begin{equation}
\alpha \sim f(\alpha)
\end{equation}

\begin{equation}
f(\alpha) = \frac{1}{b-a}, a < \alpha < b
\end{equation}

\begin{equation}
f(\alpha) = 0, else, 
\end{equation}
where $a = 0$ and $b = {\alpha}_{max}$.

With this masking strategy, the distortion-aware prior is introduced into the training of the auto-encoder. Hence, the learned latent space features are forced to be in a high-level coherent structure, in which different parts of features are correlated to each other. Thus, the missing parts of the learned representation can be inferred from the rest of the existing features, which denotes \textit{the anti-distortion property}.

\subsubsection{Low-level acoustic representation}
To extract high-level acoustic representation with anti-distortion properties for speech synthesis tasks, the low-level acoustic representation should at least contain enough local acoustic texture of speech. The mel-spectrogram is a common choice for neural vocoders training. The neural vocoders, conditioned on ground truth mel-spectrogram, are capable of synthesizing very natural speech samples. Based on this previous work, we pick mel-spectrogram as the low-level acoustic representation for the speech frames reconstruction objective. Thus, the auto-encoder is trained to reconstruct mel-spectrogram with latent space features randomly masked.

\begin{figure}[!t] 
    \centering 
    \includegraphics[width=\linewidth]{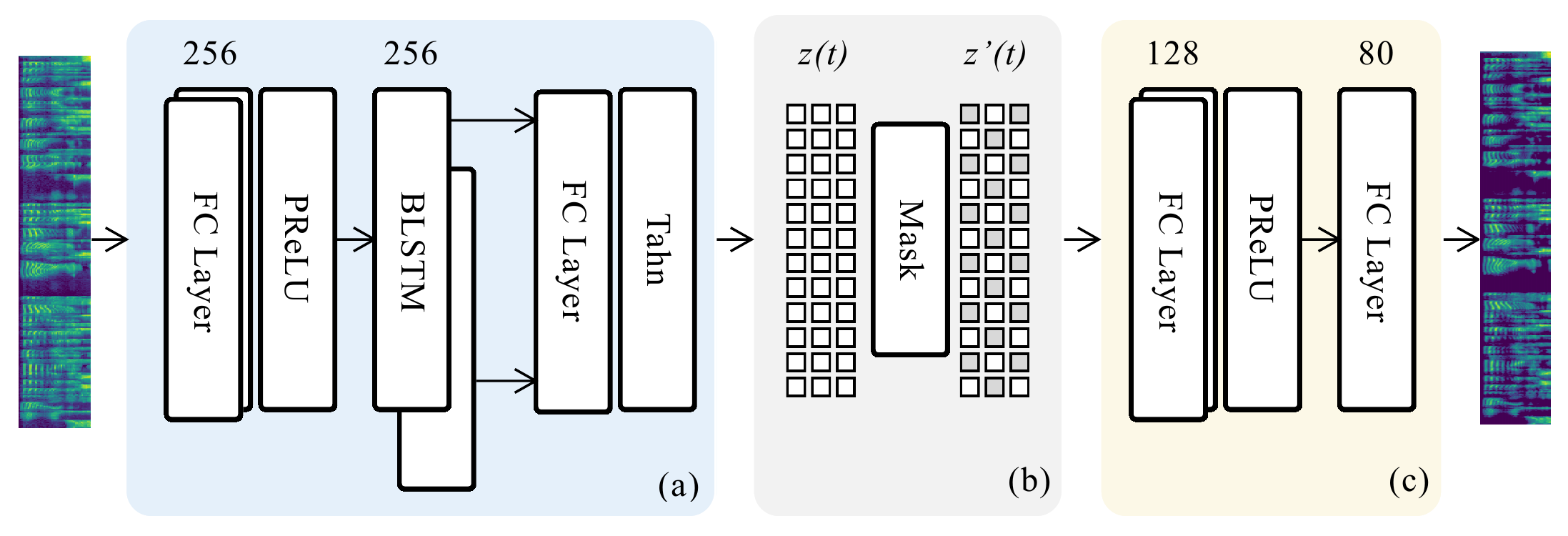}
    \caption{The detailed architecture of the auto-encoder. The numbers above the blocks represent the layer size of the network. FC Layer represents Fully-Connected Layer, PReLU represents Parametric Rectified Linear Unit, BLSTM represents Bi-directional Long-short term memory model. (a) Encoder of the auto-encoder,  (b) Masking strategy during the auto-encoder training, (c) Decoder of the auto-encoder.}
    \label{imag_ae} 
\end{figure}
\subsubsection{Using auto-encoder for representation learning}
\label{subsubsec:auto-encoder}

This masking process is the key to learning a high-level coherent structure, in which the missing parts can be inferred from the rest features. Auto-encoder is utilized for reconstructing mel-spectrogram in our proposed self-supervised pre-training with distortion-aware prior. The overall structure of auto-encoder pre-training is illustrated in Figure \ref{imag1} (a), which is a typical encoder-decoder structure but with latent space features randomly masked denoted by the white blocks after applying the masking strategy. $L_{recon}$ represents the reconstruction loss in Equation (\ref{eqa9}) and (\ref{eqa10}). In Figure \ref{imag_ae}, we present detailed structure of the auto-encoder.

To incorporate both the speech texture of the current frame and contextual information, a recurrent encoder $E(\cdot)$ is used to encode a sequence of mel-spectrogram frames $m(t)$ in bidirectional order, where $t$ is the frame time step. Since what we propose is a pre-training framework, the recurrent encoder can be of different types of layers that can capture contextual information. In our experiments, as shown in the Figure \ref{imag_ae}, the recurrent encoder is simply a stack of fully connected (FC) layers $F
_{n}(\cdot)$ and Bidirectional Long Short-Term Memory (BLSTM) layers $B_{n}(\cdot)$, followed by a \textit{tanh} activated FC layer $F(\cdot)$, where $n$ represent different number of layers. The process of encoding procedure in the auto-encoder model can be written as follows:

\begin{equation}
    E(\cdot) = F(B_{n}(F_{n}(\cdot)))
\end{equation}
\begin{equation}
    z(t) = E(m(t))
\end{equation}

For each time step $t$, the output of recurrent encoder $z(t)$ is randomly masked by masking ratio $\alpha$, which is simply implemented by dropout $drop(\cdot, \alpha)$. This is the key to learning a high-level coherent structure, in which the missing parts can be inferred from the rest features.

\begin{equation}
    z'(t) = drop(z(t), \alpha)
\end{equation}

The masked latent space features $z'(t)$ is then fed into an decoder $D(\cdot)$ to output reconstructed mel-spectrogram frames $\hat{m}(t)$. The decoder is designed to be simple and only consists of several FC layers solely for feed-forward feature mapping. 

\begin{equation}
    D(\cdot) = F_{n}(\cdot)
\end{equation}

\begin{equation}
    \hat{m}(t) = D(z'(t))
\end{equation}

This design forces the learned high-level coherent structure of speech to have mainly existed in the latent space features, which makes these learned features as informative as possible. Overall, as illustrated in Figure \ref{imag1} (a), the pre-training stage follows a common auto-encoder training scheme with Mean-Square Error (MSE) as a reconstruction criterion for better reconstruction of the acoustic features. 

\begin{equation}
    \min\limits_{E(\cdot), D(\cdot)} L_{recon} \label{eqa9}
\end{equation}
where
\begin{equation}
    L_{recon} = \mathbb{E}[||\hat{m}(t) - m(t)||_2^2] \label{eqa10}
\end{equation}

\subsection{Downstream neural vocoder training}
In common neural vocoder training~\cite{prenger_2018_Waveglow}, acoustic features like mel-spectrogram are extracted from ground-truth speech samples and condition neural vocoders for speech generation. For the downstream neural vocoder training, there are two differences compared to the common scheme. Firstly, the recurrent encoder part $E(\cdot)$ of the trained auto-encoder is jointly fine-tuned with the neural vocoder, so that the recurrent encoder can be further adapted directly to the speech samples generation. Secondly, the mel-spectrogram is replaced by the learned representation SAR to condition the neural vocoder $V(\cdot)$.

As presented in Figure \ref{imag1} (b), the mel-spectrogram of each frame $m(t)$ is firstly obtained by human-crafted signal processing from ground-truth speech samples as in a common scheme. Then, the frames of mel-spectrogram $m(t)$ are fed into the recurrent encoder $E(\cdot)$ to get the learned representation of each frame $z(t)$. The masking strategy in the pre-training stage is kept to maintain the anti-distortion property, which transforms $z(t)$ into the masked representation $z'(t)$. 

\begin{equation}
    z(t) = E(m(t))
\end{equation}
where $E(\cdot)$ is pre-trained in the section \ref{subsubsec:auto-encoder} but fine-tuned in the second stage, "Neural vocoder training", of training.

Finally, the frames of masked representation $z'(t)$ condition the neural vocoder for downstream training. The masking strategy is only activated during the training to incorporate distortion-aware prior into both the recurrent encoder and the neural vocoder. In the copy-synthesis inference of the downstream neural vocoder, the masking strategy is deactivated. In other words, frames of $z(t)$ instead of $z'(t)$ are used to condition the neural vocoder to generate speech samples. Because the fine-tuned encoder will generate self-fixed latent representations, which will replace the role of mel-spectrogram to supervise the neural vocoder training.

Hence, during the training, the conditional features are the masked learned representation $z'(t)$. During the copy-synthesis of the neural vocoder, the presentation encoder is fixed to become a feature extractor, which extracts $z(t)$ for each frame to condition the neural vocoder for waveform generation. In the fine-tuning stage, the pre-trained auto-encoder is connected with a pre-trained neural vocoder originally based on mel-spectrogram to further adapt the learned representation directly to the waveform. Finally, the learned latent space features $z(t)$ are used to replace the mel-spectrogram. 

\begin{figure}[!htbp] 
    \centering 
    \includegraphics[width=\linewidth]{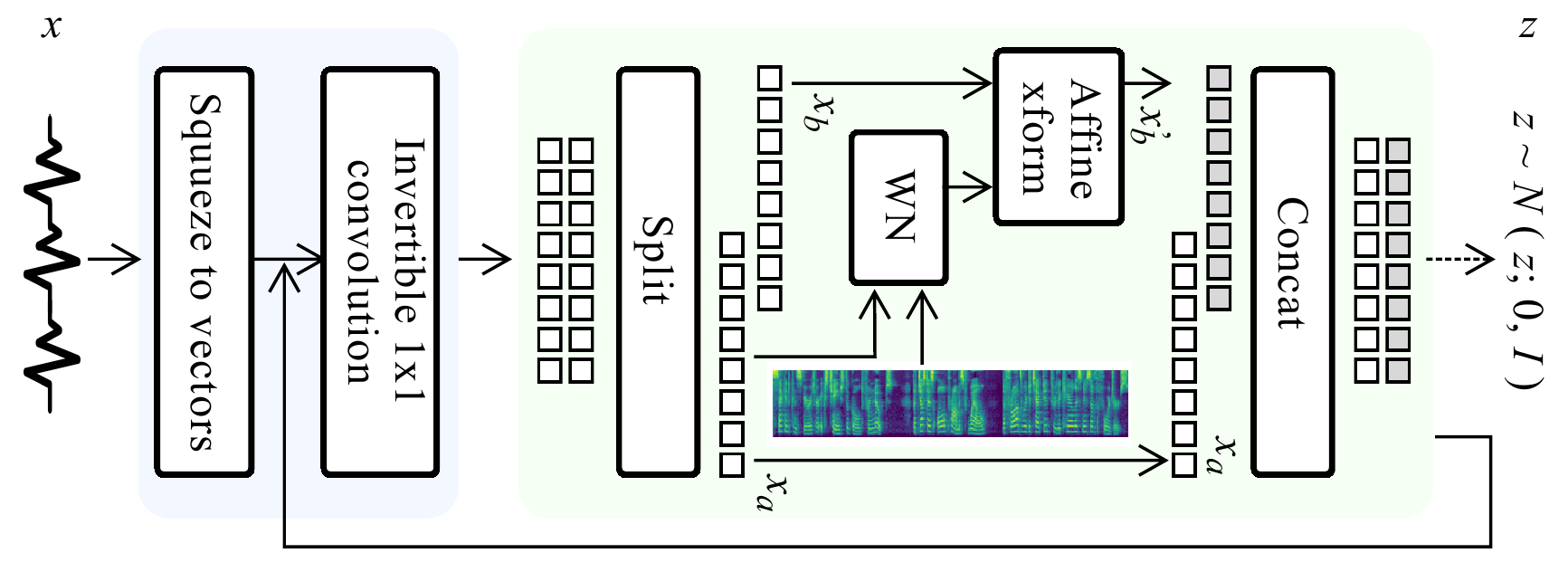}
    \caption{The architecture of neural vocoder used in downstream neural vocoder training. "Split" denotes split in the channel dimension. "WN" denotes the modified WaveNet. $x$ denotes the ground-truth audio, $z$ denotes variable sampled from the Gaussian distribution.}
    \label{imag_waveglow} 
\end{figure}

The neural vocoder in this training stage basically follows the design of Waveglow shown in the Figure \ref{imag_waveglow}, the likelihood function of the model is as follows.
\begin{equation}
\begin{aligned}
    \log p_{\theta}(x) &= - \frac{z(x)^Tz(x)}{2\sigma^2}\\ &+  \sum_{j=0}^{n_{cp}} \log s_j(x,m(t))\\ &+ \sum_{k=0}^{n_{cv}} \log \det|W_k|
\end{aligned}
\end{equation}
where
\begin{equation}
    z\sim \mathcal{N}(z;0, \bm{I})
\end{equation}

\begin{equation}
    x = f_0 \circ f_1 \circ \cdots \circ f_k(z)
\end{equation}

\begin{equation}
    z = f_k^{-1} \circ f_{k-1}^{-1} \circ \cdots \circ f_0^{-1}(x)
\end{equation} 
where the first component is derived from the log-likelihood of a spherical Gaussian distribution of the variable $z(x)$. $\sigma^2$ represents the hypothesized variance of the Gaussian distribution, while the remaining terms are included to accommodate the change of variables. $n_{cp}$ denotes number of coupling layers and $n_{cv}$ denotes number of convolutional layers. $s_j$ and $t$ denotes the change of variables when doing affine xform function. $m(t)$ denotes the mel spectrograms. $W_k$ denotes the weights used in the 1x1 convolutions. These weights are initialized to be orthonormal, thereby ensuring their invertibility. $\log \det$ denotes the log-determinant of the Jacobian function.

As illustrated in Figure \ref{imag1} (c), unlike the downstream neural vocoder training stage which jointly trains both the recurrent encoder and the neural vocoder, the downstream acoustic model training only utilizes the recurrent encoder as a feature extractor. Moreover, The masking strategy is deactivated and the acoustic model directly uses $z(t)$ instead of masked features $z'(t)$ as the target output for training. There are two conjecture about the advantage of the neural vocoder training stage. Firstly, connecting the auto-encoder with the neural vocoder further adapts the learned representation $z(t)$ directly to waveform generation, which leads to a better match. Secondly, it makes it much easier for training by connecting the trained auto-encoder and the trained neural vocoder for joint modeling, since the output of the autoencoder exactly matches the input acoustic features condition of the neural vocoder. During the inference, only the decoder part of the auto-encoder which is simply a stack of FC layers is kept connected with the neural vocoder to generate speech samples. In other words, the input latent space features $z(t)$ with the decoder part of the auto-encoder as a whole, replacing the conditional acoustic features part.

The whole inference pipeline of the back-end of the downstream-trained TTS system works just as in the common scheme. The acoustic model $A(\cdot)$ follows Tacotron-2 which accepts text information as input $s$ and generates a sequence of predicted acoustic representation $\hat{z}(t)$. Then, the frames of $\hat{z}(t)$ condition the neural vocoder to generate speech samples. The objective function of the model is as follows:

\begin{equation}
    \min\limits_{A(\cdot)}  L = \mathbb{E}[||\hat{z}(t) - z(t)||_2^2]
\end{equation}
where
\begin{equation}
    \hat{z}(t) = A(s)
\end{equation}

\begin{equation}
    z(t) = E(m(t))
\end{equation}

\subsection{Inference stage}
During the inference, this whole back-end of the TTS system uses the learned acoustic representation. Firstly the predicted representation $\hat{z}(t)$ is generated by the acoustic model, then the $\hat{z}(t)$ conditions the neural vocoder to generate speech samples. 

\begin{equation}
    \hat{z}(t) = A(m(t))
\end{equation}

\begin{equation}
    \hat{x} = V(\hat{z}(t))
\end{equation}

Thus, for the acoustic model, the only difference is the change of target acoustic features. Although the new predicted acoustic features $\hat{z}$ still have a gap with their ground truth counterpart, they have anti-distortion properties to significantly reduce the effects of distortions.

\section{Experiments}

\subsection{Dataset}
We evaluate our proposed representation on both English and Mandarin datasets. For English, the self-supervised pre-training is based on VCTK~\cite{vctk} dataset (109 English speakers with 400 sentences each speaker) and the downstream neural vocoder is trained with LJSPEECH~\cite{ljspeech17} dataset which is a single-speaker dataset with 13,100 sentences. The Mandarin dataset used for pre-training is an internal dataset, which is composed of 3 males and 3 females with an average of around 9,000 read-out sentences for each speaker. 
% The internal Mandarin corpus is recorded in a small room with walls attached with sound-absorbing materials. 
The downstream Mandarin dataset is a single-speaker CSMSC~\cite{csmsc} corpus with 10,000 sentences in total. The above datasets are split into training, validation, and test sets at percentages of $90\%$, $5\%$, and $5\%$ respectively.

% The objective evaluation is conducted at a $16kHz$ sampling rate, so all the waveform files from the above datasets that have a higher sampling rate are downsampled. 

\subsection{Model configuration}
In our experiments of the auto-encoder self-supervised pre-training, the encoder of the auto-encoder structure first transforms an 80-dim mel-spectrum through two FC layers with 256 hidden units. To circumvent the issue of the output of the ReLU activation function being entirely zero when all input features are negative, the Parametric Rectified Linear Unit (PReLU) was implemented as a nonlinear activation function. After the FC transform, two BLSTM layers are stacked to encode the whole sequence in a many-to-many scheme, with the number of output states being 256 as well. Finally, an FC layer further transforms the output of BLSTM into $z(t)$ with a Hyperbolic tangent (Tanh) activation function to constrain the representation to be within $[-1, 1]$. The decoder is much simpler by stacking a PReLU activated FC layer and a linear FC layer to output the final acoustic representation. The first FC layer contains 128 hidden units, whereas the output FC layer consists of 80 units. As for training, the batch size used is 64, and Adam Optimizer is applied with 1e-4 learning rate. For checkpoint selection, an early stop is applied regarding the validation loss. 
% The ${\alpha}_{max}$ is set to $0.2$, so that 
The $\alpha$ is sampled from a uniform distribution of interval $[0, 0.2]$.

For downstream tasks in our experiments, Waveglow is selected as a neural vocoder. We use similar parameters setup to those in~\cite{prenger_2018_Waveglow}. 
% 12 coupling layers and 12 invertible 1x1 convolutions are used, with each coupling layer having 8 layers of dilated convolutions. 512 channels are used as residual connections and 256 channels are used in the skip connections. 
Each audio clip contains 16,000 samples. The batch size used is 8 with Adam Optimizer of learning rate 1e-4. For ground-truth acoustic feature extraction, Fast Fourier Transform (FFT) window size is selected to be 1024 with a hop size equal to 16 milliseconds (256 for 16000 sampling rate, and 128 for 8000 sampling rate). As for the acoustic model, Tacotron2~\cite{shen_2017_natural} is utilized to predict acoustic features. The Pre-net dimension is set to 256, and the number of embedding dimensions is selected to be 512 with a dimension of attention recurrent neural network (RNN) equal to 1024. The dimension of decoder RNN is set to 1024 with the dimension of Postnet embedding equal to 512. The rest of the hyper-parameters are the same with~\cite{shen_2017_natural}.

\subsection{Evaluation systems}
To objectively compare the anti-distortion property of the learned representation under different corruption conditions, we built 3 copy-synthesis systems based on both Mandarin and English datasets:
\begin{itemize}
    \item Mel-WaveGlow: Waveglow based on mel-spectrogram.
    \item SAR-WaveGlow: Waveglow based on the learned representation with auto-encoder self-supervised pre-training.
    \item AR-WaveGlow: Waveglow based on the learned representation without auto-encoder self-supervised pre-training, i.e., the recurrent encoder of the auto-encoder is directly connected to a trained Waveglow for downstream fine-tune training.
\end{itemize}

We designed two types of distortions to corrupt the extracted ground-truth acoustic features:
\begin{itemize}
    \item White noise: white noise is simulated by sampling from a normal Gaussian distribution, which is additive noise directly added to the acoustic features with a target Signal-to-Noise Ratio (SNR). In our experiments setup, we tried two different target SNRs: 10 $dB$ and 15 $dB$.
    \item Masking: Similar to the masking strategy in the auto-encoder pre-training, we randomly masked the acoustic features with different masking ratios $\alpha=0.1$ and $\alpha=0.2$.
\end{itemize}

% \subsection{Evaluation metrics}
% Extended Short-Time Objective Intelligibility (ESTOI)~\cite{jensen_2016_an} is commonly used for the intelligibility assessment of speech synthesis technologies and is calculated by measuring the distortion in spectrotemporal modulation patterns of the generated speech signal. Therefore, the ESTOI score is sensitive to both inaccurate reconstructions of the spectral profile and the inconsistencies in the reconstructed temporal patterns. Thus, by computing the ESTOI scores between the generated waveform of different systems and the ground-truth waveform, to some extent, we can measure the anti-distortion property of the acoustic features that different systems are based on. A higher ESTOI score indicates better voice quality, and also demonstrates the stronger anti-distortion property. In our experiments, 100 sentences were randomly selected from test sets of CSMSC and LJSPEECH corpus respectively. The ESTOI scores between the groundtruth waveform and the copy-synthesis ones with two types of corruptions were computed. In the subjective evaluation, Mean Opinion Score (MOS) is chosen as the metric. 

\subsection{Objective evaluation results}

\begin{table}[htbp]
\centering
\caption{ESTOI scores comparison on the anti-distortion property on Mandarin dataset}
\label{table1}
\begin{tabular}{cccccc} 
\toprule
\multirow{2}{*}{Distortion type} & \multirow{2}{*}{Raw} & \multicolumn{2}{c}{Mask ratio}  &   \multicolumn{2}{c}{SNR} \\ \cline{3-6} 
& & $\alpha = 0.1$ & $\alpha = 0.2$ & 15 $dB$ & 10 $dB$  \\ \midrule
Mel-WaveGlow & \textbf{0.927} & 0.837 & 0.754 & 0.830 & 0.726    \\ 
AR-WaveGlow  & 0.881 & 0.862 & 0.847 & 0.859 & 0.822 \\ \hline
SAR-WaveGlow & 0.891 & \textbf{0.877} & \textbf{0.855} & \textbf{0.881} & \textbf{0.859} \\
\bottomrule
\end{tabular}
\end{table}

\begin{table}[htbp]
\centering
\caption{ESTOI scores comparison on the anti-distortion property on English dataset}
\label{table2}
\begin{tabular}{cccccc} 
\toprule
\multirow{2}{*}{Distortion type} & \multirow{2}{*}{Raw} & \multicolumn{2}{c}{Mask ratio}  &   \multicolumn{2}{c}{SNR} \\ \cline{3-6} 
& & $\alpha = 0.1 $& $\alpha = 0.2 $& 15 $dB$ & 10 $dB$  \\ \midrule
Mel-WaveGlow &  \textbf{0.904} & 0.783 & 0.681 & 0.782 & 0.667   \\ 
AR-WaveGlow  & 0.859 & 0.841 & 0.816 & 0.856 & 0.842 \\ \hline
SAR-WaveGlow & 0.866 & \textbf{0.855} & \textbf{0.830} & \textbf{0.860} &\textbf{0.846} \\
\bottomrule
\end{tabular}
\end{table}

The Extended Short-Time Objective Intelligibility (ESTOI) scores are computed for objective evaluation. The ESTOI score is sensitive to both to both incorrect spectral profile reconstructions and inconsistent temporal pattern reconstructions. A higher ESTOI score indicates better voice quality and demonstrates a stronger anti-distortion property.
Tables \ref{table1} and \ref{table2} describe the ESTOI evaluation on 100 sentences randomly selected from the test sets of the CSMSC and LJSPEECH corpora, respectively. For the uncorrupted acoustic features, the original mel-spectrogram achieves the best ESTOI score. However, in cases of different distortion conditions, the ESTOI scores of Mel-WaveGlow degraded dramatically. While the ESTOI scores of SAR-WaveGlow in those conditions suffer minor loss compared to the raw condition. This shows the strong anti-distortion property of the learned representation against both types of corruption. Even the noise-adding corruption is not seen during the representation learning, the learned presentation still generalizes well. Also, by comparing the ESTOI scores of SAR-WaveGlow and AR-WaveGlow, the self-supervised pre-training part of our proposal proved to be necessary for anti-distortion representation training. Because for both Table \ref{table1}, every ESTOI score in SAR-WaveGlow is dominantly better than that in AR-WaveGlow.

\begin{figure}[htbp]
    \centering
    \subfigure[Waveglow copy-synthesis]{
        \includegraphics[width=0.46\linewidth]{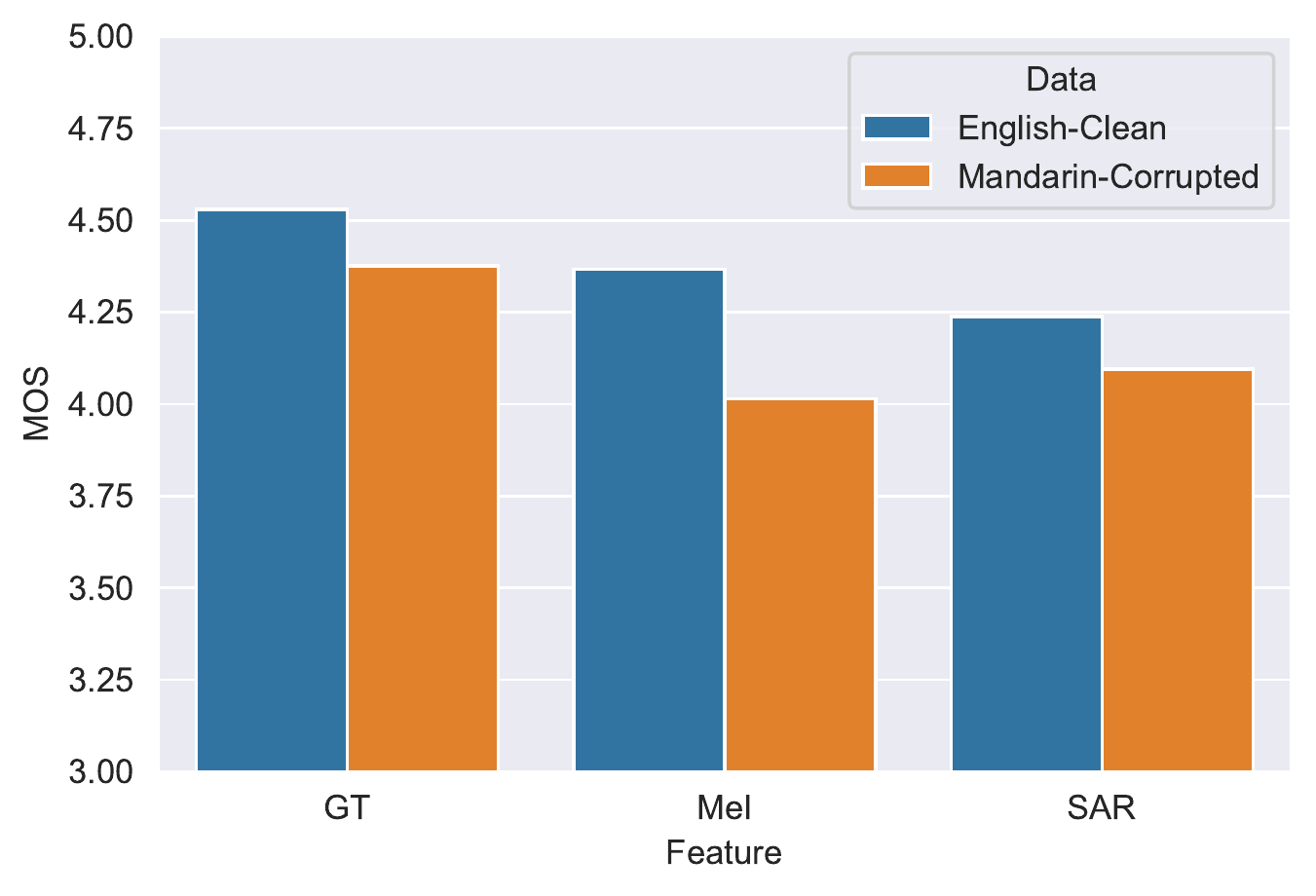}
\label{subfig:a}
    }
    \hspace{0mm}
    \subfigure[TTS system]{
        \includegraphics[width=0.46\linewidth]{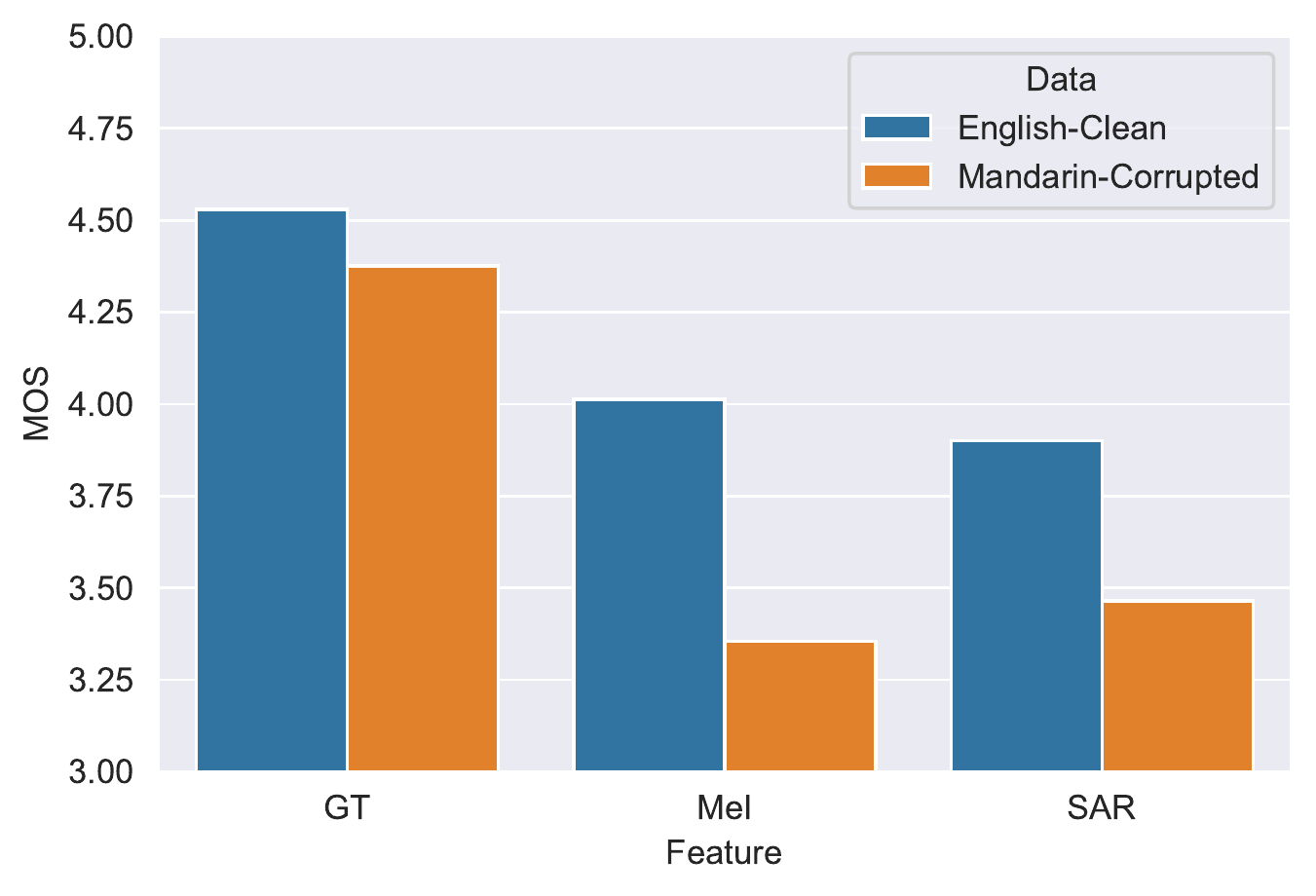}
\label{subfig:b}
    }
    \caption{MOS test results of different tasks on uncorrupted or corrupted datasets}
      \label{fig:02}
    \end{figure}
    
\subsection{Subjective evaluation results}

To evaluate the anti-distortion property of the learned representation in speech synthesis scenarios, we built Mandarin TTS systems for subjective evaluation. According to our presumption, there exists a coherent structure within acoustic features that help recover missing parts. And in real-world scenarios, there are some cases in that we may only be able to retrieve low-sampling-rate speech data of a specific speaker due to limited internet bandwidth or device limits, and yet we still want to build a high-sampling-rate TTS system. To mimic those scenarios, we simply simulate a low-quality speech corpus by firstly downsampling and then upsampling the speech data. Specifically, we downsampled the CSMSC corpus from 16000 $Hz$ to 8000 $Hz$, and then upsampled it back to 16000 $Hz$, so that the high-frequency information is deleted. Based on this corrupted CSMSC corpus, we separately trained two Tacotron2 models, between which the only difference is the acoustic features, i.e., mel-spectrogram $m(t)$ versus the learned representation $z(t)$. The previously trained 16000 $Hz$ Waveglow models were reused for both acoustic features. 

We did two Mean Opinion Score (MOS) evaluations based on the same setup: 50 sentences were selected from CSMSC test set and 30 Mandarin speakers participated in score ratings. We first did a MOS test on results from the copy-synthesis of Waveglow models using different acoustic features.  As shown in the Figure \ref{subfig:a}, 
% the results are consistent with our objective evaluation in Table \ref{table1} that 
for uncorrupted acoustic features the neural vocoder based on mel-spectrogram achieves the best voice quality. The learned representation introduces minor degradation of the copy-synthesis voice quality, compared to the mel-spectrogram. The other MOS test was conducted to compare the anti-distortion property of different predicted acoustic features from acoustic models based on low-quality corrupted CSMSC corpus. Although the learned representation introduces degradation into the copy-synthesis waveform, according to Figure \ref{subfig:b}, the learned representation shows stronger robustness to the low-quality training data for acoustic modeling. The learned representation together with the jointly trained neural vocoder seems to be capable of recovering some of the missing high-frequency information in the generated speech.

\section{Conclusions}
In this work, we propose an anti-distortion self-supervised learning framework to create a new acoustic representation (SAR) to replace the handcrafted mel-spectrogram. This is based on the intuition that the mel-spectrogram contains a high-level coherent structure. We can reconstruct the missing parts from the rest of the features. By introducing distortion-aware prior to the auto-encoder pre-training stage, the anti-distortion property is granted to SAR. This anti-distortion property is verified by both objective and subjective analyses. We proved that the self-supervised pre-training stage is necessary for learning a representation with the anti-distortion property. Moreover, the anti-distortion property of SAR is superior to that of the mel-spectrogram, and it also generalizes to unseen corruptions like white noise addition. We also built TTS systems based on SAR, and the subjective analysis of which shows the robustness of SAR on low-quality training data.

% Although SAR shows strong anti-distortion properties according to our experiments, it also introduces some downsides to voice quality, especially for copy-synthesis without corruption of acoustic features. Besides, the TTS system using SAR built with low-quality training data still suffers large voice quality degradation compared to those built with high-quality training data, even though it shows better robustness compared to the mel-spectrogram. This could be partly due to the loss of some local texture of speech after the encoding process of the auto-encoder. To address this issue, a better framework can be designed to emphasize local information or turn to a neural vocoder with more generative capabilities like generative adversarial neural vocoders. The distortion-aware self-supervised learning grants the learned representation with anti-distortion property, which is verified from both objective and subjective perspectives. Since what we proposed is a framework, the model structures of the auto-encoder can be replaced for customized purposes. For example, the recurrent encoder part of the auto-encoder structure used in our experiments can be replaced by a more powerful structure like Wavenet. Also, the copy-synthesis degradation issue may be alleviated by replacing Waveglow with a GAN-based vocoder.

\section{Ackowledgement}
Supported by the Key Research and Development Program of Guangdong Province (grant No. 2021B0101400003). Corresponding author is Xulong Zhang (zhangxulong@ieee.org).

\bibliographystyle{IEEEtran}
\bibliography{DSAA01}
\end{document}